\documentclass[12pt]{article}
\usepackage{graphicx}
\newcommand{\thspace}{\kern.08333em}

\def \bea{\begin{eqnarray}}
\def \beq{\begin{equation}}

\def \cn{Collaboration}

\def \eea{\end{eqnarray}}
\def \eeq{\end{equation}}
\def \gb{\bar \Gamma}
\def \gs{\stackrel{>}{\sim}}

\def \ls{\stackrel{<}{\sim}}

\def \s{\sqrt{2}}

\def \v#1#2{V_{#1#2}}

\begin{document}

\rightline{EFI 03-34}
\rightline{hep-ph/0307095}

\bigskip
\bigskip
\centerline{\bf RATES AND ASYMMETRIES IN $B \to K \pi$ DECAYS
\footnote{To be published in Physics Letters B.}}

\bigskip
\centerline{Michael Gronau\footnote{Permanent Address: Physics
Department,
Technion -- Israel Institute of Technology, 32000 Haifa, Israel.}
and Jonathan L. Rosner}
\centerline{\it Enrico Fermi Institute and Department of Physics}
\centerline{\it University of Chicago, Chicago, IL 60637}
\vskip 1cm

\centerline{\bf ABSTRACT}
\bigskip

\begin{quote}
We discuss a potential discrepancy in an approximate relation among $B \to K
\pi$ rates which, with increased statistical significance, would imply new 
physics in $\Delta I=1$ transitions.  An approximate relation between 
CP-violating rate differences in $B^0/\overline{B}^0 \to K^{\pm}\pi^{\mp}$ 
and $B^{\pm} \to K^{\pm}\pi^0$ is used to combine these rate differences
to reduce upper limits on the two CP asymmetries. These rates and 
asymmetries are used to update bounds  on the CKM phase $\gamma$.
\end{quote}
\bigskip

\leftline{\qquad PACS codes:  12.15.Hh, 12.15.Ji, 13.25.Hw, 14.40.Nd}
\bigskip

In Ref.\ \cite{comb} we proposed 
separate relations among decay rates and among direct CP asymmetries in $B
\to K \pi$ decays, following from a model-independent hierarchy among various
contributions to decay amplitudes.  At that time three of these decays, $B^0
\to K^+\pi^-,~B^+\to K^+\pi^0$ and $B^+ \to K^0\pi^+$, had been observed, while
a fourth, $B^0 \to K^0 \pi^0$, still remained to be seen.  The question of
direct CP asymmetries in these decays remained very much an open one. We noted
the conditions under which one expected the following sum rule
to hold \cite{comb,Lipkin}:
\beq \label{eqn:sumrule}
2 \Gamma(B^+ \to K^+ \pi^0) + 2 \Gamma(B^0 \to K^0 \pi^0)
\approx \Gamma(B^+ \to K^0 \pi^+) + \Gamma(B^0 \to K^+ \pi^-)~~~,
\eeq
and derived an approximate relation between the rate differences in the decays
involving $K^+$:
\bea \label{eqn:combCP}
\Delta( K^+ \pi^-) & \equiv & \Gamma(\bar B^0 \to K^- \pi^+) -
\Gamma(B^0 \to K^+ \pi^-) \simeq \nonumber \\ 2 \Delta(K^+ \pi^0) & \equiv &
2[\Gamma(B^- \to K^- \pi^0) - \Gamma(B^+ \to K^+ \pi^0)]~~~
\eea
which would allow one to combine such rate differences to improve the
statistical accuracy of either one.  In the present note we update these
analyses, as well as one \cite{GR02} in which decay rates and asymmetries are
combined in order to obtain limits on phases of the Cabibbo-Kobayashi-Maskawa
(CKM) matrix.  (Preliminary accounts of some of this last work have appeared in
Refs.\ \cite{conf}.)

The decay rates and CP asymmetries which we use are summarized in Table
\ref{tab:data}.  We use averages of CLEO \cite{CLbr}, BaBar \cite{Babr},
and Belle \cite{Bebr} measurements compiled in Ref.\ \cite{CGR}, and new BaBar
results \cite{Ocariz} on $B^+ \to K^0 \pi^+$.  To relate
branching ratios to decay rates we have used $\tau^+ = (1.656 \pm 0.014)$ ps
and $\tau^0 = (1.539 \pm 0.014)$ ps \cite{LEPBOSC} for the respective $B^+$
and $B^0$ lifetimes.  CP asymmetries are defined as
\beq
A_{CP}(f) = \frac{\Gamma(\bar B \to \bar f) - \Gamma(B \to f)}
{\Gamma(\bar B \to \bar f) +  \Gamma(B \to f)}~~~.
\eeq

\begin{table}
\caption{CP-averaged branching ratios, CP-averaged decay rates, and CP rate
asymmetries for $B \to K \pi$ decays.  Branching ratios and CP asymmetries are
based on averages in Ref.\ \cite{CGR} except for $B^+ \to K^0 \pi^+$, where
we have used new BaBar results \cite{Ocariz} in our averages.
\label{tab:data}}
\begin{center}
\begin{tabular}{c c c c} \hline \hline
       Mode         &    Branching     &     Partial     &   $A_{CP}$   \\ 
                    & ratio ($10^{-6}$) & width ($10^{-9}$ eV) &        \\
\hline
$B^0 \to K^+ \pi^-$ & $18.16 \pm 0.79$ & $7.77 \pm 0.35$ & $-0.088\pm0.040$ \\
$B^0 \to K^0 \pi^0$ & $11.21 \pm 1.36$ & $4.79 \pm 0.58$ &              \\
$B^+ \to K^0 \pi^+$ & $20.62 \pm 1.35$ & $8.19 \pm 0.54$ & $~~0.003\pm0.059$ \\
$B^+ \to K^+ \pi^0$ & $12.82 \pm 1.07$ & $5.10 \pm 0.43$ & $~~0.035\pm0.071$ \\
\hline \hline
\end{tabular}
\end{center}
\end{table}

The sum rule (\ref{eqn:sumrule}) is based on the following amplitude
decomposition \cite{hier}:
\bea
A(B^0 \to K^+ \pi^-)    & = & -(p+t)~~~, \nonumber \\
\s A(B^0 \to K^0 \pi^0) & = & p-c~~~, \nonumber \\
A(B^+ \to K^0 \pi^+)    & = & p~~~, \nonumber \\
\s A(B^+ \to K^+ \pi^0) & = & -(p+t+c)~~~,
\eea
\beq
t \equiv T + P_{\rm EW}^C~,~~ c \equiv C + P_{\rm EW}~,~~
p \equiv P - \frac{1}{3} P_{\rm EW}^C~~~.
\eeq
The terms $T$ and $C$ represent color-favored and color-suppressed tree
amplitudes while $P$ stands for a (gluonic) penguin amplitude.  Color-favored
and color-suppressed electroweak penguin amplitudes are represented by
$P_{\rm EW}$ and $P_{\rm EW}^C$.  Small annihilation and exchange
amplitudes are neglected.  These amplitudes obey a hierarchy \cite{hier}
\beq\label{hierarchy}
|P| \gg |T|, |P_{\rm EW}| \gg |C|, |P^C_{\rm EW}|~~.
\eeq
We find
\beq \label{eqn:piz}
2|A(B^0 \to K^0 \pi^0)|^2 + 2|A(B^+ \to K^+ \pi^0)|^2 =
2|p|^2 + 2{\rm~Re} (p^*t) + |t|^2 + 2{\rm~Re}(c^*t) + 2|c|^2,
\eeq
\beq
\label{eqn:pich}
|A(B^0 \to K^+ \pi^-)|^2 + |A(B^+ \to K^0 \pi^+)|^2 =
2|p|^2 + 2{\rm~Re}(p^*t) + |t|^2~~~.~~~~~~~~~~~~~~~~~~\null
\eeq 
The sum rule (\ref{eqn:sumrule}) follows from neglecting 
in Eq.\ (\ref{eqn:piz}) the last two terms which are second order in 
the small ratios $|t/p|$ and $|c/p|$.
It holds separately for the decays shown and their
CP-conjugates. In Ref.\ \cite{CGR} the last two terms 
are found to contribute at most $4\%$ of the dominant $2|p|^2$ terms.

Using the experimental values for CP-averaged partial widths in Table
\ref{tab:data}, the sum rule reads
\beq \label{eqn:compar}
(19.8 \pm 1.4) \times 10^{-9}~{\rm eV} =
(16.0 \pm 0.6) \times 10^{-9}~{\rm eV}~~~.
\eeq
The left-hand side differs from the right-hand side by $(3.8 \pm 1.6) \times
10^{-9}$ eV, or $(24 \pm 10)$\% of the better-known right-hand side. This
discrepancy is too large to be
accounted for by the neglected standard-model terms. If it is not caused by
new physics effects, the most likely source is a systematic underestimate of
the efficiency for $\pi^0$ detection in each experiment.

An enhancement of $B\to K\pi$ modes involving a 
neutral pion would be interpreted as a new physics amplitude in $\Delta I =1$ 
$B\to K\pi$ transitions. Written in terms of isospin amplitudes, the sum rule 
(\ref{eqn:sumrule}) reads \cite{comb} 
\beq
|B_{1/2}|^2 +|A_{1/2} -2A_{3/2}|^2 \approx
|B_{1/2}|^2 +|A_{1/2} + A_{3/2}|^2~~,
\eeq
where $A$ and $B$ are $\Delta I =1$ and $\Delta I =0$ amplitudes 
and subscripts denote the isospin of $K\pi$.  They are related to $p$, $c$,
and $t$ by \cite{comb}
\beq \label{eqn:ispin}
B_{1/2} = p + \frac{t}{2}~~,~~~A_{1/2} = \frac{2c-t}{6}~~,~~~A_{3/2} =
- \frac{c+t}{3}~~~.
\eeq
The sum rule holds when
\beq
  3|\,|A_{3/2}|^2 - 2\,{\rm Re}(A_{1/2}A^*_{3/2})| \ll |B_{1/2}|^2\,~~.
\eeq 
In the Standard Model the left-hand-side is given  by the terms 
${\rm Re}(c^*t) + |c|^2$ in Eq.~(\ref{eqn:piz}), where $c$ is dominated
by $P_{\rm EW}$. These pure $\Delta I=1$ terms were estimated to be at most 
about $4\%$ of the right-hand-side.  Current data favor a larger positive value
for these terms.  A significant discrepancy in the sum rule beyond a few
percent (which could be associated with isospin violations stemming from
$m_u \ne m_d$)
would imply $\Delta I =1$ contributions from physics beyond the Standard 
Model. Models involving such amplitudes and several other manifestations in 
$B\to K\pi$ decays were studied in \cite{GKN}.

The validity or violation of the sum rule affects interpretations of various
$B \to K \pi$ rate ratios \cite{GR02,GRKpi,NR,BF,Matias}.  Three rate 
ratios provide useful information on weak phases, especially when combined 
with information on CP asymmetries:
\bea
R & \equiv & \frac{\gb(B^0 \to K^+ \pi^-)}{\gb(B^+ \to K^0 \pi^+)}
= 0.948 \pm 0.074~~, \label{eqn:R0} \\
R_c & \equiv & \frac{2 \gb(B^+ \to K^+ \pi^0)}{\gb(B^+ \to K^0 \pi^+)}
= 1.24 \pm 0.13~~, \label{eqn:Rc} \\
R_n & \equiv & \frac{\gb(B^0 \to K^+ \pi^-)}{2 \gb(B^0 \to K^0 \pi^0)}
= 0.81 \pm 0.10~~, \label{eqn:R0p}
\eea
where we have used the averages of Table \ref{tab:data}, and $\gb$ stands
for a CP-averaged partial width.
                                          
To first order in terms of order $|t/p|$ and $|c/p|$ (where $c$ is dominated by
$P_{\rm EW}$), the ratios (\ref{eqn:Rc}) and (\ref{eqn:R0p}) should be equal.  
In fact, at this order the equality of the two ratios of rates holds 
separately for $B^+$ and $B^0$ and for $B^-$ and $\bar B^0$. To see this,
we write
\beq
\frac{2\Gamma(B^+ \to K^+\pi^0)}{\Gamma(B^+ \to K^0\pi^+)}  = 
\left| \frac{p+c+t}{p} \right|^2~~,~~~
\frac{\Gamma(B^0 \to K^+\pi^-)}{2\Gamma(B^0 \to K^0\pi^0)} = 
\left| \frac{p+t}{p-c} \right|^2~~,~~~
\eeq
and use the binomial expansion for $(p-c)^{-1} = p^{-1}(1 - \frac{c}{p})^{-1}$
in the second relation.  
Alternatively, we can show that $R_c = R_n$ to this order by writing
\bea
2 \gb(B^+ \to K^+ \pi^0) & = & \gb(B^+ \to K^0 \pi^+)(1 + \epsilon_{+0})~~,
\nonumber \\
\gb(B^0 \to K^+ \pi^-) & = & \gb(B^+ \to K^0 \pi^+)(1 + \epsilon_{+-})~~,
\nonumber \\
2 \gb(B^0 \to K^0 \pi^0) & = & \gb(B^+ \to K^0 \pi^+)(1 + \epsilon_{00})~~.
\eea
The sum rule (\ref{eqn:sumrule}) implies $\epsilon_{+0} + \epsilon_{00} =
\epsilon_{+-}$, or to first order in small quantities $\epsilon$,
\beq
R_c  = (1 + \epsilon_{+0}) = \frac{1 + \epsilon_{+-}}{1 + \epsilon_{00}}
= R_n~~~.
\eeq
The fact that $R_c$ and $R_n$ differ so much \cite{Yoshikawa}, being nearly 
$2 \sigma$ above and below 1, respectively, is directly related to the
large violation of the sum rule (\ref{eqn:sumrule}).  Thus, until the source
of the sum rule violation is clarified, one should view results based on
either ratio with some caution.
We shall show below that one may cancel out effects of imperfectly determined
$\pi^0$ detection efficiency by considering the quantity $(R_c R_n)^{1/2}$. 

We now turn to the relation (\ref{eqn:combCP}).  
At a leading order in $|T/P|,~|P_{\rm EW}/P|$ and $|C/T|$, the two rate 
differences are equal \cite{comb}, since they involve 
a common interference term of $p$ and $t$ (namely $P$ and $T$). The rate 
difference $\Delta(K^+\pi^0)$ contains also a higher order interference of 
$p$ and $c$ (namely, $P$ and $C$) dominating $\Delta(K^0\pi^0)$, and an 
even higher order interference of $P_{\rm EW}$ and $C$.
Using the partial widths in Table \ref{tab:data}, we find
\beq                      |
\Delta(K^+ \pi^-) = (-0.67 \pm 0.31) \times 10^{-9}~{\rm eV}~~,~~~           
2 \Delta(K^+ \pi^0) = (0.36 \pm 0.71) \times 10^{-9}~{\rm eV}~~~.
\eeq
These two partial rate differences are consistent with each other and
with zero.  If constrained to be equal and averaged, they give
\beq \label{eqn:avdelta}
\Delta(K^+ \pi^-) = 2 \Delta(K^+ \pi^0) = (-0.52 \pm 0.29) \times 10^{-9}~{\rm
eV}~~.
\eeq
The $K^+ \pi^-$ asymmetry clearly carries more weight.  The implied CP
asymmetries are then
\beq 
A_{CP}(B^0 \to K^+ \pi^-) = -0.066 \pm 0.037~~,~~~
A_{CP}(B^+ \to K^+ \pi^0) = -0.051 \pm 0.028~~~.
\eeq
These can be used, if desired, in updated analyses along the lines of Ref.\
\cite{GR02}, to interpret experimental ranges of $R$ and $R_c$ in terms
of limits on the weak CKM phase $\gamma$.  Instead, we shall use the
observed CP asymmetries separately in each channel, since, as we shall show,
neither analysis is very sensitive to small changes in the CP asymmetries
as long as these are already small.

\begin{figure}
\begin{center}
\includegraphics[height=5.5in]{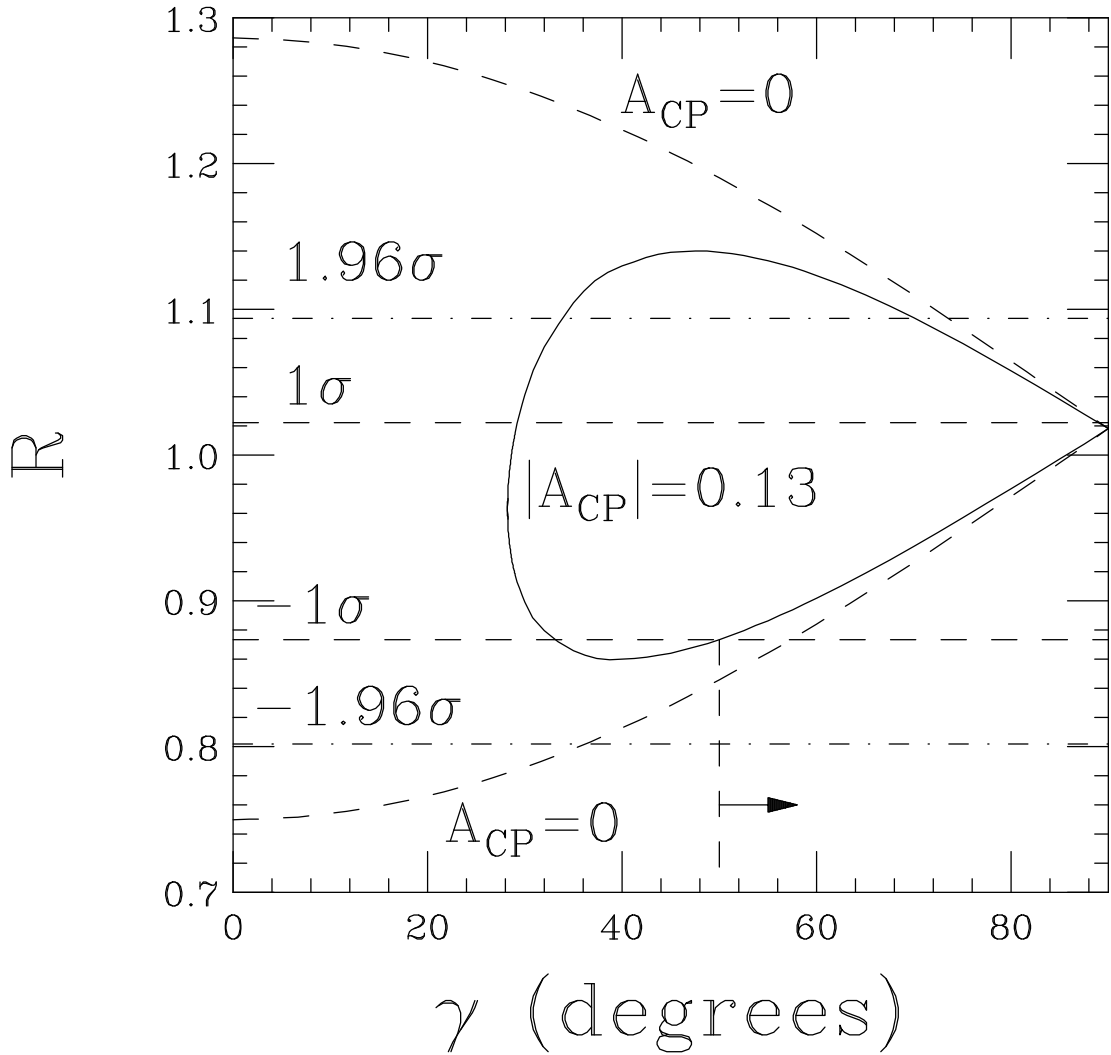}
\caption{Behavior of $R$ for $r = 0.13$ and $A_{CP}(B^0 \to K^+ \pi^-) = 0$
(dashed curves) or $|A_{CP}(B^0 \to K^+ \pi^-)| = 0.13$ (solid curve) as a
function of the weak phase $\gamma$.
Horizontal dashed lines denote $\pm 1 \sigma$ experimental limits on $R$,
while dot-dashed lines denote $95\%$ c.l. ($\pm 1.96 \sigma$) limits.
The upper branches of the curves correspond to the case $\cos \gamma
\cos \delta <0$, while the lower branches correspond to $\cos \gamma
\cos \delta >0$.
\label{fig:Racp}}
\end{center}
\end{figure}

The decay $B^+ \to K^0 \pi^+$ is a pure penguin ($p$) process, while the
amplitude for $B^0 \to K^+ \pi^-$ is proportional to $p + t$, where $t$ is a
tree amplitude.  The ratio $t/p$ has magnitude $r$, weak phase $\gamma \pm
\pi$ (depending on convention), and strong phase $\delta$.  The ratio 
$R$ of these two rates (averaged
over a process and its CP conjugate) is
\beq \label{eqn:Rval}
R = 1 - 2 r \cos \gamma \cos \delta + r^2~~~,
\eeq
The CP asymmetry in $B^0 \to K^+ \pi^-$ is
$A_{CP}(B^0 \to K^+ \pi^-) = - 2 r (\sin \gamma \sin \delta)/R$.
One may eliminate $\delta$ between this equation and Eq.\ (\ref{eqn:Rval})
and plot $R$ as a function of $\gamma$ for the allowed range of
$A_{CP}(B^0 \to K^+ \pi^-)$.  The value of $r$, based on present branching
ratios and arguments given in Refs.\ \cite{GR02,GRKpi}, is $r=0.17 \pm 0.04$.
The average in Table \ref{tab:data} implies $|A_{CP}(B^0 \to K^+ \pi^-)| \le
0.13$ at the $1 \sigma$ level.  Curves for $A_{CP}(B^0 \to K^+ \pi^-) =0$ 
and $|A_{CP}(B^0 \to K^+ \pi^-)| = 0.13$
are shown in Fig.\ \ref{fig:Racp}. The lower limit $r = 0.13$ is used to
generate these curves since the limit on $\gamma$ will be the most 
conservative.

At the $1 \sigma$ level, using the constraints that $R$ must lie between 0.873
and 1.022 and $|A_{CP}(B^0 \to K^+ \pi^-)|$ must lie between zero and 0.13, 
one finds $\gamma \gs 50^\circ$.  (We consider only those values of $\gamma$
allowed at 95\% confidence level by fits to other observables \cite{CKMf},
$38^\circ \le \gamma \le 80^\circ$.  Thus although values of $\gamma \ls
31^\circ$ are allowed in Fig.\ \ref{fig:Racp}, we do not consider them
further.  We adopt a similar restriction for other bounds to be presented
below.)  No bound can be obtained at the 95\% confidence level,
however.  If one were to use the improved bound on $A_{CP}(B^0 \to K^+ \pi^-)$
implied by Eq.\ (\ref{eqn:combCP}), a slight improvement on the $1 \sigma$
lower bound on $\gamma$ would result.  Reduction of errors on $R$
and improvement of the estimate of $r$ would have a much greater impact.

The comparison of rates for $B^+ \to K^+ \pi^0$ and $B^+ \to K^0 \pi^+$ gives
similar information on $\gamma$.  The amplitude for $B^+ \to K^+ \pi^0$ is
proportional to $p + t + c$, where $c$ contains a color-suppressed amplitude.
Originally it was suggested that this amplitude be compared with $p$ from $B^+
\to K^0 \pi^+$ and $t+c$ taken from $B^+ \to \pi^+ \pi^0$ using flavor SU(3)
\cite{GRL} using a triangle construction to determine $\gamma$.  However,
electroweak penguin (EWP) amplitudes contribute significantly in the $t+c$ 
term \cite{EWP}. It was noted subsequently \cite{NR} that since the combination
$t+c$ corresponds to isospin $I(K \pi) = 3/2$ for the final state [see
Eq.\ (\ref{eqn:ispin})], the strong-interaction phase of its EWP contribution
is the same as that of the rest of the $t+c$ amplitude and the ratio of the two
contributions is given in terms of known Wilson coefficients and CKM factors.
This permits a calculation of the EWP correction.

\begin{figure}
\begin{center}
\includegraphics[height=5.5in]{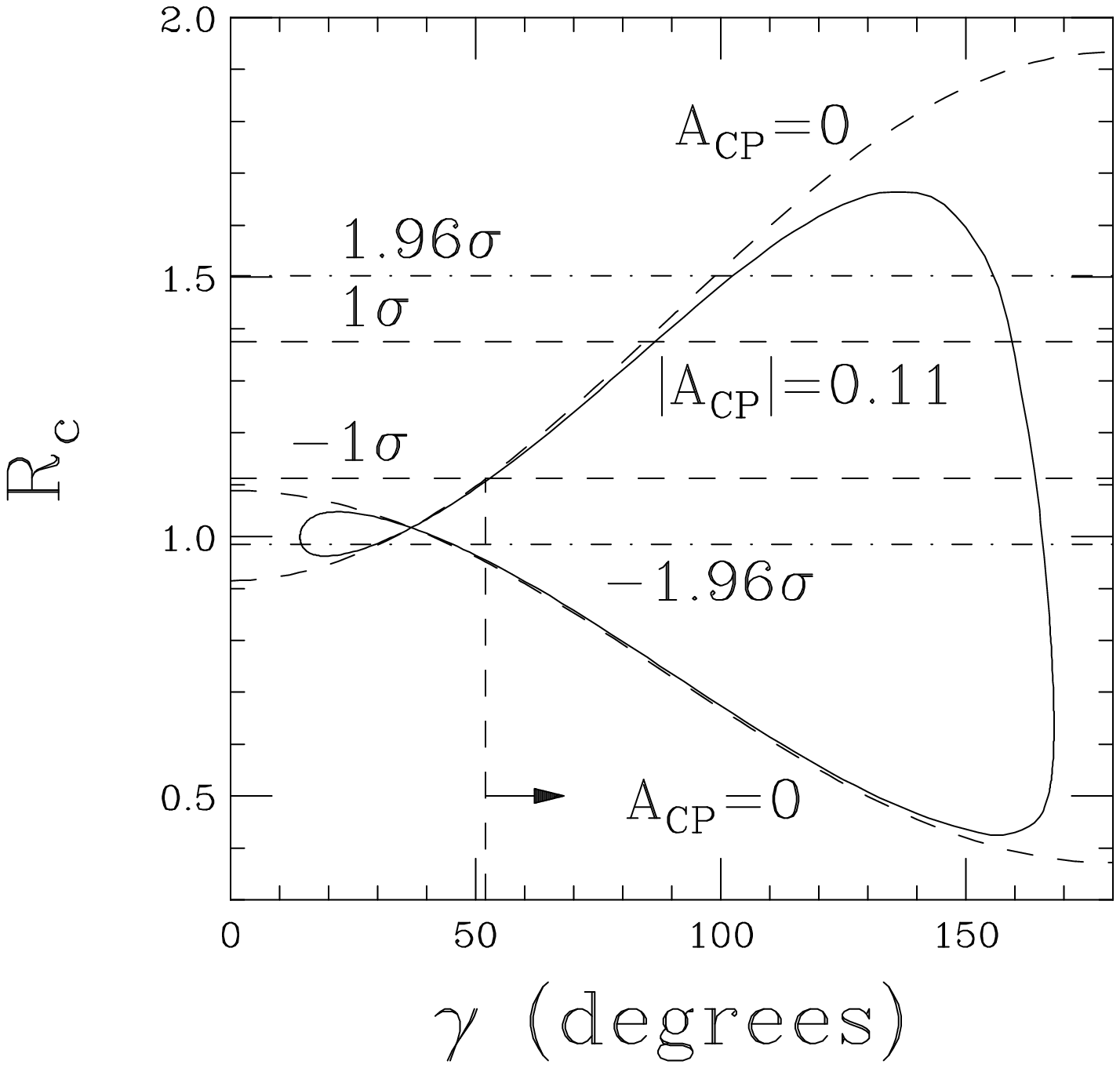}
\caption{Behavior of $R_c$ for $r_c = 0.22$ ($1 \sigma$ upper limit) and
$A_{CP}(K^+ \pi^0) = 0$ (dashed curves) or $|{\cal A}_{CP}(K^+ \pi^0)|
= 0.11$ (solid curve) as a function of the
weak phase $\gamma$. Horizontal dashed lines denote $\pm 1 \sigma$ experimental
limits on $R_c$, while dotdashed lines denote 95\% c.l. ($ \pm 1.96 \sigma$)
limits.  Upper branches of curves correspond to $\cos \delta_c(\cos \gamma -
\delta_{EW}) < 0$, while lower branches
correspond to $\cos \delta_c(\cos \gamma - \delta_{EW}) > 0$.  Here we have
taken $\delta_{EW} = 0.80$ (its $1 \sigma$ upper limit), which
leads to the most conservative bound on $\gamma$.
\label{fig:Rcacp}}
\end{center}
\end{figure}

New data on branching ratios and CP asymmetries permit an update of previous
analyses \cite{GR02,NR}.  The expressions for the rate ratio and CP asymmetry
are
\beq \label{eqn:Rce}
R_c  =  1 - 2 r_c \cos \delta_c~(\cos \gamma - \delta_{\rm EW}) 
 +  r_c^2(1 - 2 \delta_{EW} \cos \gamma + \delta_{EW}^2)~~,
\eeq
\beq \label{eqn:Accp}
A_{CP}(K^+ \pi^0) =  - 2 r_c \sin \delta_c \sin \gamma/ R_c~~,
\eeq
where $r_c \equiv |(T+C)/P| = 0.20 \pm 0.02$, and $\delta_c$ is a strong
phase, eliminated by combining (\ref{eqn:Rce}) and (\ref{eqn:Accp}).
One must also use an estimate \cite{NR} of the electroweak penguin parameter
$\delta_{\rm EW} = 0.65 \pm 0.15$.  One obtains the most conservative (i.e.,
weakest) bound on $\gamma$ for the maximum values of $r_c$ and $\delta_{\rm
EW}$ \cite{GR02}.  The resulting plot is shown in Fig.\ \ref{fig:Rcacp}.  One
obtains a bound at the $1 \sigma$ level very similar to that in the previous
case:  $\gamma \gs 52^\circ$.  The bound is set by the curve for
{\it zero} CP asymmetry, as emphasized in Ref.\ \cite{NR}.  Consequently,
the improved estimate of $A_{CP}(B^+ \to K^+ \pi^0)$ has no impact on this
bound.

If the deviations from unity of $R_c$ and $R_n$ are a consequence of an
underestimate of the efficiency for $\pi^0$ detection, one may compensate
for this effect by considering their geometric mean:
\beq
(R_c R_n)^{1/2} = \left[ \frac{\gb(B^+ \to K^+ \pi^0)}{\gb(B^+ \to K^0 \pi^+)}
\frac{\gb(B^0 \to K^+ \pi^-)}{\gb(B^0 \to K^0 \pi^0)} \right]^{1/2}
= 1.004 \pm 0.084~~.
\eeq
Since one partial width for a decay involving a $\pi^0$ appears in the
numerator while another appears in the denominator, the efficiencies will
cancel one another.  Since we have argued that to first order in small
quantities $R_c$ and $R_n$ should be equal, this ratio should also be 
given in this approximtion by
Eq.~(\ref{eqn:Rc}), and should provide an equally valid limit on $\gamma$.
The neglect of second order terms, as well as of rescattering effects,
probably amounts to corrections of a few percent in $R$, $R_c$, and $R_n$,
and hence of a few degrees in $\gamma$.

\begin{figure}
\begin{center}
\includegraphics[height=5.5in]{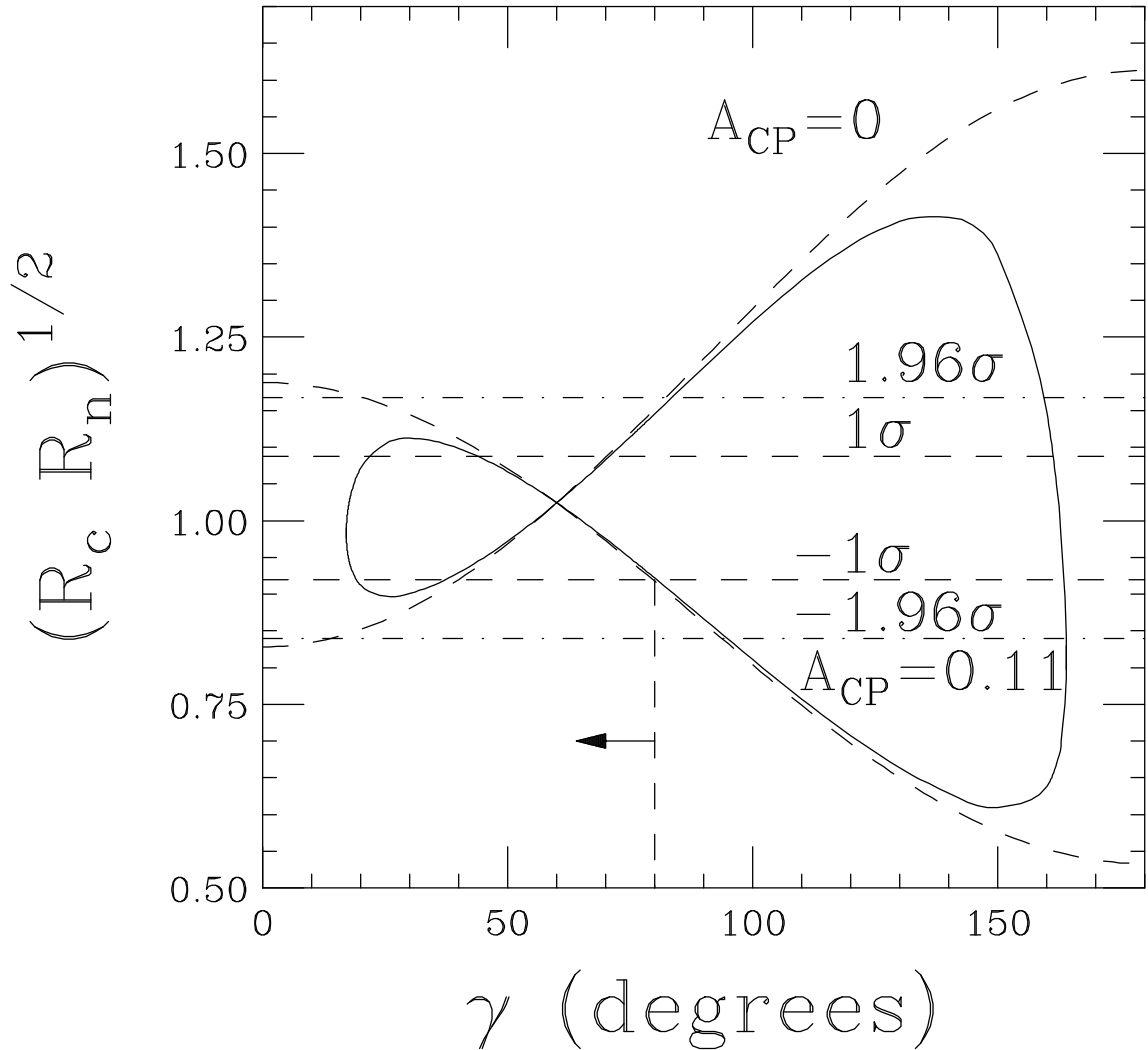}
\caption{Behavior of $(R_c R_n)^{1/2}$ for $r_c = 0.18$ ($1 \sigma$ lower
limit) and ${\cal A}_{CP}(K^+ \pi^0) = 0$ (dashed curves) or
$|{\cal A}_{CP}(K^+ \pi^0)| = 0.11$ (solid curve) as a function of the
weak phase $\gamma$. Horizontal dashed lines denote $\pm 1 \sigma$ experimental
limits on $(R_c R_n)^{1/2}$, while dotdashed lines denote 95\% c.l. ($ \pm 1.96
\sigma$) limits.  Upper branches of curves correspond to $\cos \delta_c(\cos
\gamma - \delta_{EW}) < 0$, while lower branches
correspond to $\cos \delta_c(\cos \gamma - \delta_{EW}) > 0$.  Here we have
taken $\delta_{EW} = 0.50$ (its $1 \sigma$ lower limit), which
leads to the most conservative bound on $\gamma$.
\label{fig:Rmean}}
\end{center}
\end{figure}

Since $(R_c R_n)^{1/2}$ is so close to unity, it turns out that the most
conservative bound occurs for the {\it smallest} values of $r_c$ and
$\delta_{EW}$, respectively 0.18 and 0.50, and for $|A_{CP}(B^+ \to K^+ \pi^0)|$
at its upper limit of 0.11. The resulting plot is shown in Fig.\
\ref{fig:Rmean}.
One obtains an {\it upper} limit in this case:  $\gamma \ls 80^\circ$ at the
$1 \sigma$ level.  The $1 \sigma$ limits $50^\circ \ls \gamma \ls 80^\circ$,
obtained from $R$ and $(R_c R_n)^{1/2}$ and from the CP asymmetries in 
$B^0\to K^+\pi^-$ and $B^+\to K^+\pi^0$ 
are to be compared with those from a global fit to CKM parameters
\cite{CKMf}:  $44^\circ \le \gamma \le 72^\circ$ at 68\% c.l.\ or
$38^\circ \le \gamma \le 80^\circ$ at 95\% c.l.

We comment further on what would be required to enhance $\gb(B^+ \to K^+
\pi^0)$ and $\gb(B^0 \to K^0 \pi^0)$ by ${\cal O}(25\%)$, leading to the
observed deviations of $R_c$ and $R_n$ from unity.  It is sufficient to take
a suitable linear combination of the $\Delta I = 1$ amplitudes $A_{1/2}$
and $A_{3/2}$ such that only neutral-pion emission, and not charged-pion
emission, is affected.  This corresponds to an amplitude transforming as
$c$ in Eqs.\ (4) and (5). The new amplitude cannot enhance both  
decay rates by interfering with the dominant $\Delta I = 0$ amplitude
$B_{1/2}$ (which is the only one receiving a contribution from the dominant
penguin term $p$), since the interference terms are of opposite sign.  One can
see this from the relative $p$ and $c$ contributions in Eqs.\ (4).  The new
amplitude has to be nearly half in magnitude and $90^\circ$ out of phase with
respect to $B_{1/2}$ so that its absolute square would give
the needed enhancement for both $B \to K \pi^0$ decays.

We imagine two types of operators.  The first, transforming as an electroweak
penguin $P_{EW}$ contributing to the $c$ amplitude in Eq.\ (5), could
contribute to $\bar b \to \bar s \pi^0$.  One would expect the corresponding
$\bar b \to \bar d \pi^0$ amplitude to be suppressed by a factor of $|V_{cd}/
V_{us}| \simeq 0.23$ and hence to have little effect in $\Delta S = 0$
$B$ decays.  On the other hand, if there were a term transforming as the
$C$ part of $c$ in Eq.\ (5), for example due to a serious mis-estimate of
a rescattering contribution to the color-suppressed amplitude, one would
expect the $\Delta S = 0$ process to be {\it enhanced} by a factor of
$|V_{ud}/V_{us}|$ with respect to the $|\Delta S| = 1$ contribution.  Such
an enhanced color-suppressed amplitude would certainly have been noticed in
$B^+ \to \pi^+ \pi^0$ and $B^0 \to \pi^0 \pi^0$, and can be ruled out.

In conclusion, the $B \to K \pi$ decay rates are approximately in the ratios 
of 2:1:2:1
expected for the $K^+ \pi^-$, $K^0 \pi^0$, $K^0 \pi^+$, and $K^+ \pi^0$ modes
if the penguin amplitude ($p$) is dominant.  However, the deviations from these
rates that one would expect due to interference with the smaller tree ($t$) 
and electroweak penguin ($P_{\rm EW}$) amplitudes do not follow the expected
pattern. Rather,
there appears to be a slight enhancement of {\it both} modes involving a
$\pi^0$ with respect to the penguin-dominance expectation.  As a result, the
sum rule (\ref{eqn:sumrule}) is poorly satisfied.  This suggests that the
use of such ratios as $R_c$ and $R_n$ to constrain CKM phases be viewed with
some caution, if the problem lies with estimates of $\pi^0$
detection efficiency.  In such a case the ratio $R$ may be more reliable.
We find that by combining it with the CP asymmetry in $B^0 \to K^+ \pi^-$
one can place a $1\sigma$ lower bound $\gamma \gs 50^\circ$. The 
corresponding $1\sigma$ bound obtained by considering $R_c$ is $\gamma 
\gs 52^\circ$. An upper $1\sigma$ bound $\gamma \ls 80^\circ$ is obtained 
from the geometric mean $(R_c R_n)^{1/2}$, in which neutral-pion 
detection efficiencies cancel one another.

We have shown that the relation (\ref{eqn:combCP}) between CP-violating rate
differences is satisfied, and that a modest improvement on errors in CP
asymmetries $A_{CP}$ may be obtained by assuming it to hold.  However, somewhat
surprisingly, further progress in the study of $B \to K \pi$ decays may depend
more on the resolution of the puzzle surrounding the sum rule
(\ref{eqn:sumrule}) than on more precise determinations of CP asymmetries.

If the discrepancy in the sum rule (\ref{eqn:sumrule}) persists at a
higher level of statistical significance, one would be forced to consider
its origin in physics beyond the standard model. The most likely 
interpretation of an enhancement of $B^+\to K^+\pi^0$ and $B^0\to K^0\pi^0$
would be that it originates in a new effective Hamiltonian transforming as
$\Delta I=1$.
\bigskip


We thank Yuval Grossman, Alex Kagan, Matthias Neubert, Arkady Vainshtein,
Mikhail Voloshin, and Lincoln Wolfenstein for helpful discussions.  Part of
this research was performed at the Aspen Center for Physics.
This work was supported in part by the United States Department of Energy
through contract DE FG02 90ER40560.

\def \ajp#1#2#3{Am.\ J. Phys.\ {\bf#1}, #2 (#3)}
\def \apny#1#2#3{Ann.\ Phys.\ (N.Y.) {\bf#1}, #2 (#3)}
\def \app#1#2#3{Acta Phys.\ Polonica {\bf#1}, #2 (#3)}
\def \arnps#1#2#3{Ann.\ Rev.\ Nucl.\ Part.\ Sci.\ {\bf#1}, #2 (#3)}
\def \art{and references therein}
\def \cmts#1#2#3{Comments on Nucl.\ Part.\ Phys.\ {\bf#1}, #2 (#3)}
\def \cn{Collaboration}
\def \cp89{{\it CP Violation,} edited by C. Jarlskog (World Scientific,
Singapore, 1989)}
\def \econf#1#2#3{Electronic Conference Proceedings {\bf#1}, #2 (#3)}
\def \efi{Enrico Fermi Institute Report No.}
\def \epjc#1#2#3{Eur.\ Phys.\ J.\ C {\bf#1} (#3) #2}
\def \ib{{\it ibid.}~}
\def \ibj#1#2#3{~{\bf#1} (#3) #2}
\def \ijmpa#1#2#3{Int.\ J.\ Mod.\ Phys.\ A {\bf#1}, #2 (#3)}
\def \ite{{\it et al.}}
\def \jhep#1#2#3{JHEP {\bf#1} (#3) #2}
\def \jpb#1#2#3{J.\ Phys.\ B {\bf#1}, #2 (#3)}
\def \mpla#1#2#3{Mod.\ Phys.\ Lett.\ A {\bf#1} (#3) #2}
\def \nat#1#2#3{Nature {\bf#1}, #2 (#3)}
\def \nc#1#2#3{Nuovo Cim.\ {\bf#1}, #2 (#3)}
\def \nima#1#2#3{Nucl.\ Instr.\ Meth.\ A {\bf#1}, #2 (#3)}
\def \npb#1#2#3{Nucl.\ Phys.\ B~{\bf#1}, #2 (#3)}
\def \npps#1#2#3{Nucl.\ Phys.\ Proc.\ Suppl.\ {\bf#1}, #2 (#3)}
\def \PDG{Particle Data Group, K. Hagiwara \ite, \prd{66}{010001}{2002}}
\def \pisma#1#2#3#4{Pis'ma Zh.\ Eksp.\ Teor.\ Fiz.\ {\bf#1}, #2 (#3) [JETP
Lett.\ {\bf#1}, #4 (#3)]}
\def \pl#1#2#3{Phys.\ Lett.\ {\bf#1}, #2 (#3)}
\def \pla#1#2#3{Phys.\ Lett.\ A {\bf#1}, #2 (#3)}
\def \plb#1#2#3{Phys.\ Lett.\ B {\bf#1} (#3) #2}
\def \prl#1#2#3{Phys.\ Rev.\ Lett.\ {\bf#1} (#3) #2}
\def \prd#1#2#3{Phys.\ Rev.\ D\ {\bf#1} (#3) #2}
\def \prp#1#2#3{Phys.\ Rep.\ {\bf#1} (#3) #2}
\def \ptp#1#2#3{Prog.\ Theor.\ Phys.\ {\bf#1}, #2 (#3)}
\def \rmp#1#2#3{Rev.\ Mod.\ Phys.\ {\bf#1} (#3) #2}
\def \rp#1{~~~~~\ldots\ldots{\rm rp~}{#1}~~~~~}
\def \si90{25th International Conference on High Energy Physics, Singapore,
Aug. 2-8, 1990}
\def \slc87{{\it Proceedings of the Salt Lake City Meeting} (Division of
Particles and Fields, American Physical Society, Salt Lake City, Utah, 1987),
ed. by C. DeTar and J. S. Ball (World Scientific, Singapore, 1987)}
\def \slac89{{\it Proceedings of the XIVth International Symposium on
Lepton and Photon Interactions,} Stanford, California, 1989, edited by M.
Riordan (World Scientific, Singapore, 1990)}
\def \smass82{{\it Proceedings of the 1982 DPF Summer Study on Elementary
Particle Physics and Future Facilities}, Snowmass, Colorado, edited by R.
Donaldson, R. Gustafson, and F. Paige (World Scientific, Singapore, 1982)}
\def \smass90{{\it Research Directions for the Decade} (Proceedings of the
1990 Summer Study on High Energy Physics, June 25--July 13, Snowmass,
Colorado),
edited by E. L. Berger (World Scientific, Singapore, 1992)}
\def \tasi{{\it Testing the Standard Model} (Proceedings of the 1990
Theoretical Advanced Study Institute in Elementary Particle Physics, Boulder,
Colorado, 3--27 June, 1990), edited by M. Cveti\v{c} and P. Langacker
(World Scientific, Singapore, 1991)}
\def \TASI{{\it TASI-2000:  Flavor Physics for the Millennium}, edited by
J. L. Rosner (World Scientific, 2001)}
\def \yaf#1#2#3#4{Yad.\ Fiz.\ {\bf#1}, #2 (#3) [Sov.\ J.\ Nucl.\ Phys.\
{\bf #1}, #4 (#3)]}
\def \zhetf#1#2#3#4#5#6{Zh.\ Eksp.\ Teor.\ Fiz.\ {\bf #1}, #2 (#3) [Sov.\
Phys.\ - JETP {\bf #4}, #5 (#6)]}
\def \zpc#1#2#3{Zeit.\ Phys.\ C {\bf#1}, #2 (#3)}
\def \zpd#1#2#3{Zeit.\ Phys.\ D {\bf#1}, #2 (#3)}

\end{document}